\documentclass{IEEEtran}
\usepackage{cite}
\usepackage{amsmath,amssymb,amsfonts}
\usepackage{graphicx}
\usepackage{textcomp,nicefrac}
\usepackage{url}
\usepackage{hyperref}
\usepackage{makecell}
\usepackage{tikz}

\def\BibTeX{{\rm B\kern-.05em{\sc i\kern-.025em b}\kern-.08em
T\kern-.1667em\lower.7ex\hbox{E}\kern-.125emX}}
\begin{document}
\title{Inter-detector differential fuzz testing for tamper detection in gamma spectrometers}
\author{Pei Yao Li, Jayson R.~Vavrek, Sean~Peisert, \IEEEmembership{Senior Member, IEEE}
    \thanks{P.Y.~Li and J.R.~Vavrek are with the Nuclear Science Division, Lawrence Berkeley National Laboratory, Berkeley, CA, 94720, USA. Correspondence: {\tt jvavrek@lbl.gov}.}
    \thanks{S.~Peisert is with Computing Sciences Research, Lawrence Berkeley National Laboratory, Berkeley, CA, 94720, USA, and the Computer Science and Public Health Departments, University of California, Davis, Davis, CA, 95616, USA.}
    \thanks{The project was funded by the U.S.~Department of Energy, National Nuclear Security Administration, Office of Defense Nuclear Nonproliferation Research and Development (DNN R\&D).}
}

\maketitle

\begin{abstract}
    We extend the concept of physical differential fuzz testing as an anti-tamper method for radiation detectors [Vavrek et al., \textit{Science and Global Security} (2025)] to comparisons across multiple detector units.
    The method was previously introduced as a tamper detection method for authenticating a single radiation measurement system in nuclear safeguards and treaty verification scenarios, and works by randomly sampling detector configuration parameters to produce a sequence of spectra that form a baseline signature of an untampered system.
    At a later date, after potential tamper opportunities, the same random sequence of configuration parameters is used to generate another series of spectra, which can be statistically compared against the baseline.
    Anomalies in the series of spectrum comparisons indicate changes in detector behavior, which may be due to tampering.
    One limitation of this original method is that once the detector has ``gone downrange'' and potentially has been tampered with, the original untampered baseline is fixed, and a new trusted baseline can never be established if additional fuzz tests at new parameters are required.
    In this work, we extend our anti-tamper fuzz testing concept to multiple detector units, such that the downrange detector can be compared against a ``golden copy'' detector that cannot have been tampered with, even despite normal inter-detector manufacturing variations.
    We show using three NaI detectors that this inter-detector differential fuzz testing can detect a representative tamper attempt, even when the tested and golden copy detectors are from different manufacturers and have different energy resolutions.
    In contrast to our previous work, detecting the tamper attempt requires visualizing the spectral comparison metric against the fuzzed parameter values and not just the fuzz sample number; moreover this baseline is non-linear and may necessitate anomaly detection methods more complicated than a simple threshold.
    Overall, our extension of physical differential fuzz testing to multiple detectors improves prospects for operationalization of the technique in real-world treaty verification and safeguards scenarios.
\end{abstract}

\section{Introduction}
\label{sec:introduction}
\IEEEPARstart{I}{n} nuclear arms control and safeguards, it is essential to ensure that radiation measurement equipment used in inspections has not been tampered with in order to hide treaty violations such as undeclared diversion and/or enrichment of nuclear material that could indicate weaponization activities.
We recently introduced \textit{physical differential fuzz testing} as an anti-tamper mechanism for radiation measurement equipment (and other sensors)~\cite{vavrek2025differential}; in this work, we extend this mechanism to comparisons among separate units of the same radiation detector type, such that random parameter sweeps on one detector can establish a signature capable of identifying tampering attempts on another detector.

\textit{Physical differential fuzz testing} extends the concept of \textit{differential fuzz testing}---a technique combining the use of \textit{fuzz random testing}~\cite{miller1990empirical, klees2018evaluating, miller2020relevance} and \textit{differential testing}~\cite{mckeeman1998differential} to establish and compare the signatures of two or more programs in response to identical but random inputs~\cite{yang2011finding}---from deterministic software systems to noisy cyber-physical systems comprising radiation detectors and their associated analysis software.
Physical differential fuzz testing randomly samples the ``state space'' of the radiation measurement system, collecting a series of spectra at different random parameter combinations as the signature of the system.
If the system's outputs are sufficiently different between two tests conducted at different times with the same inputs, it may be concluded that the detector has undergone some unexpected modification in the meantime, which could indicate tampering.
Physical differential fuzz testing therefore helps establish a kind of ``chain of knowledge'' that the detector has not been tampered with.

Recently, stakeholder feedback surfaced an operational drawback of testing a detector against only its own previous signature---once the detector has ``gone downrange,'' i.e, has been certified for and used in an arms control or safeguards verification application and potentially tampered with, the original baseline signature is fixed and no more baseline data can be collected.
This can be a problem for an inspector wishing to conduct additional fuzz testing not captured in the initial fuzz sequence (perhaps if some new potential attack pathway depending on a specific parameter combination were identified)---a new set of fuzz tests could be run on the downrange detector, but there would be no reference signature from before it went downrange.
In the same vein, detectors already sent downrange before the introduction of fuzz testing have no trusted baseline.
Instead, one may run the new fuzz sequence on both the downrange detector and on a ``golden copy'' reference detector in the inspector's laboratory that could not have undergone any tampering.
Such measurements present an interesting challenge for fuzz testing, as inter-detector manufacturing and wear differences may induce nontrivial changes in output behavior that are innocuous and not indicative of actual tampering.

In this work, we test whether the physical differential fuzz testing can also establish a ``chain of knowledge'' between different detector units---i.e. whether the fuzz random parameter sweeps on one detector can establish a signature that should be matched by another detector, even despite normal inter-detector differences in performance.
We show that this \textit{inter-detector differential fuzz testing} is still capable of detecting representative tampering attempts, making it a useful fallback from standard physical differential fuzz testing in potential future arms control and safeguards applications.

This paper is structured as follows: Section~\ref{sec:methods} describes a series of inter-detector fuzz testing experiments conducted using three separate NaI gamma spectrometers.
It describes the measurement setup and the calibration procedures used for the different detectors, as well as the representative tampering attempt that is to be detected by the inter-detector fuzzing.
Section~\ref{sec:results} presents experimental results, and shows that inter-detector fuzz testing is in fact capable of detecting the tamper attempt on one detector by comparing against a baseline signature established by another detector.
Section~\ref{sec:discussion} then provides some additional discussion on limitations and opportunities to address them in future work.

\section{Methods}\label{sec:methods}

\subsection{Experimental setup}
\begin{figure}[!htbp]
    \centering
    \fbox{\includegraphics[width=0.95\linewidth]{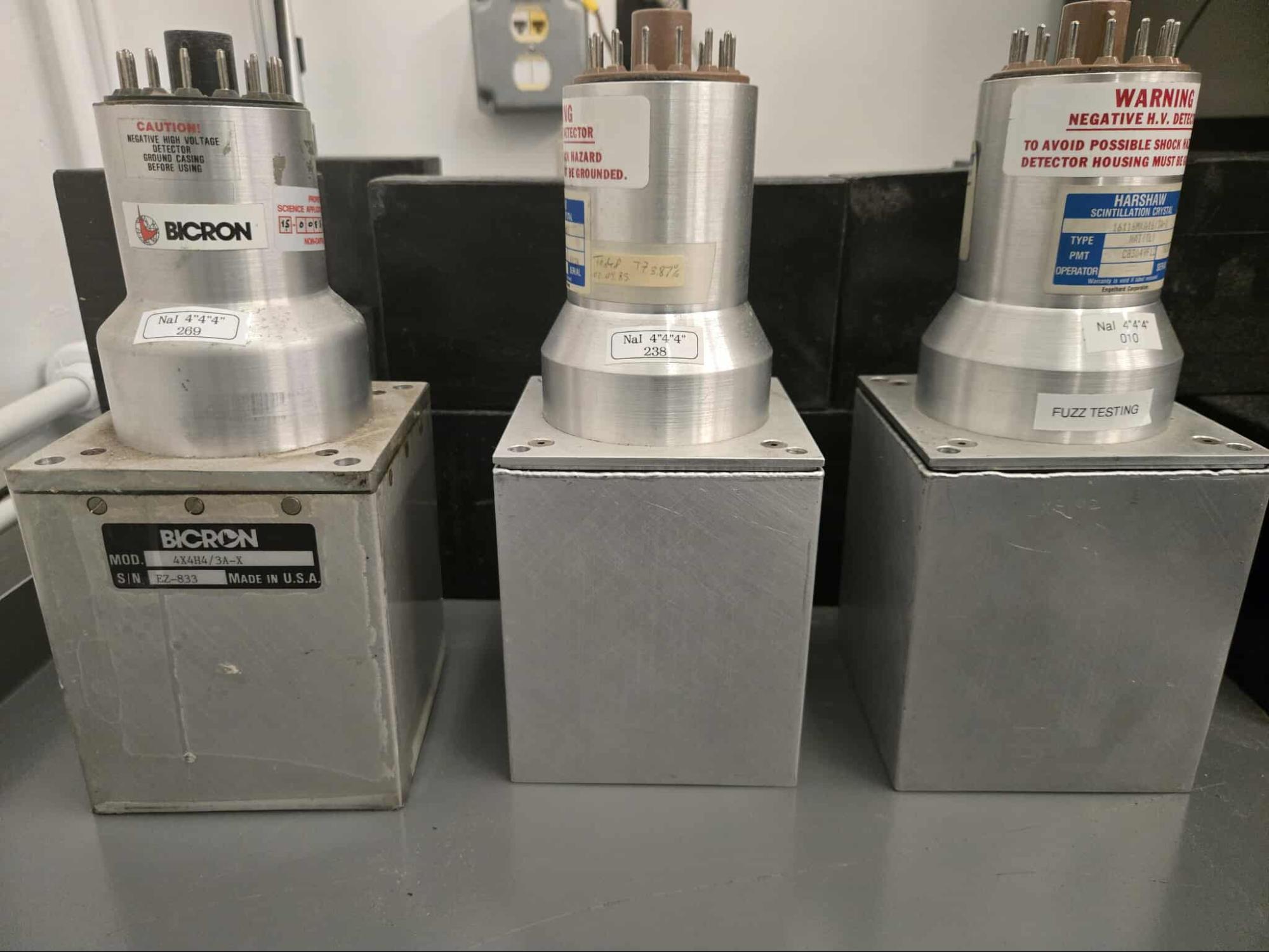}}\\
    \fbox{\includegraphics[width=0.95\linewidth]{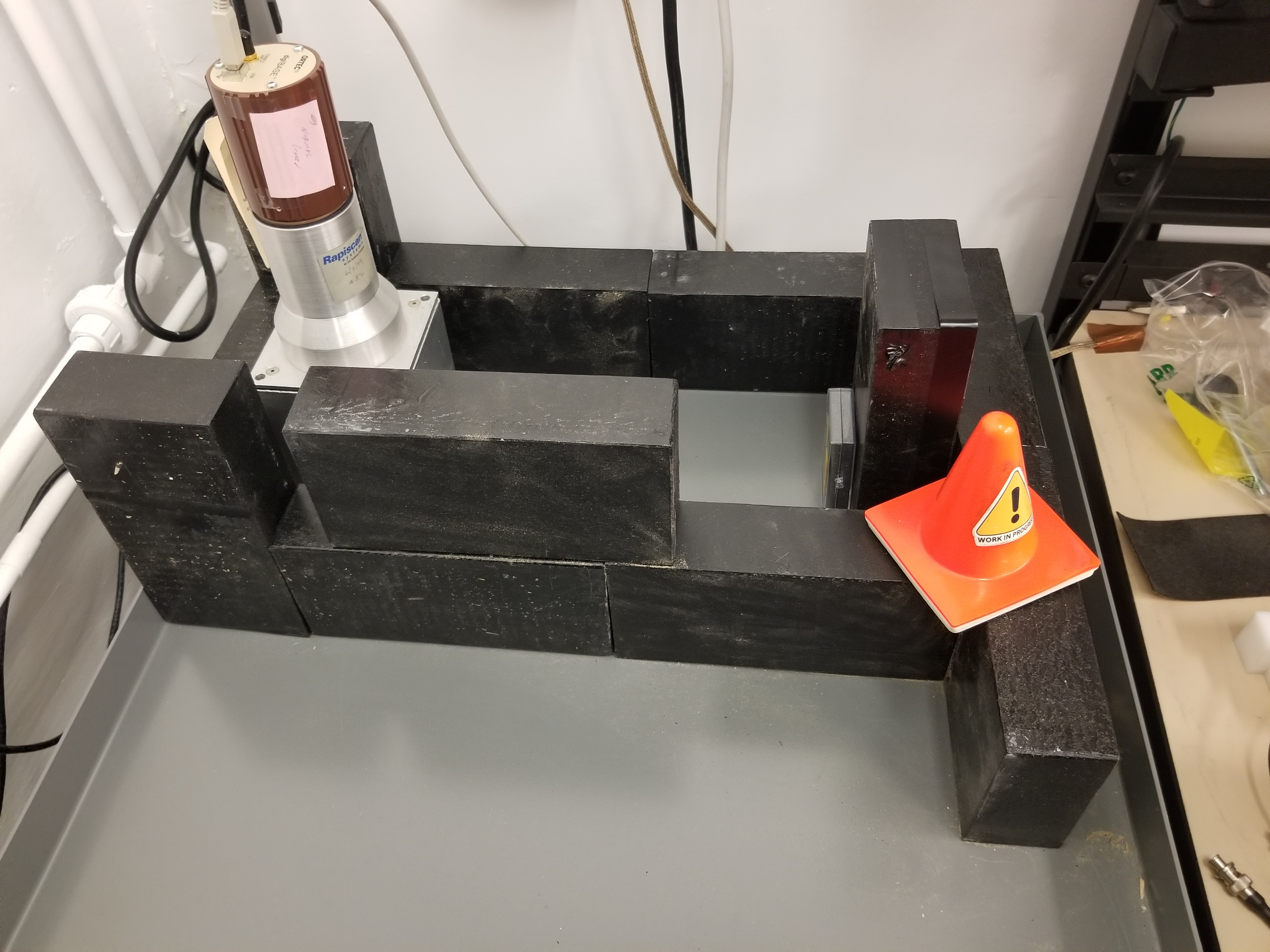}}
    \caption{
        Top: the three NaI detectors used in the inter-detector fuzzing experiments (left to right: 269, 238, 010).
        Bottom: one NaI + PMT coupled to the digitizer (brown cylinder) with a ${\sim}7$~{\textmu}Ci Cs-137 source placed about $30$~cm to the right, all within the partial lead shield.
    }
    \label{fig:setup-photo}
\end{figure}

In these experiments, we extend the fuzz testing of a single sodium iodide (NaI) radiation measurement system~\cite{vavrek2025differential} to multiple NaI detectors.
NaI detectors are used in several nuclear arms control and safeguards systems such as the Trusted Radiation Identification System (TRIS)~\cite{seager2001trusted}, the On-Line Enrichment Monitor (OLEM)~\cite{smith2011design}, and the International Atomic Energy Agency's HM-5 handheld monitor~\cite{iaea2011safeguards}.
We use three separate 4'' $\times$ 4'' $\times$ 4'' NaI gamma ray detectors (see Fig.~\ref{fig:setup-photo} and Table~\ref{tab:det-resolution}), numbered 010, 238, and 269.
We note that two of the three detectors have a different original manufacturer (Harshaw) than the third (Bicron).
Stickers on the two Harshaw detectors (010 and 238) provide measured resolutions ($\sigma$) at $662$~keV of $4.45\%$ and $3.87\%$ from $1998$ or earlier, but no historical resolution information is available for the Bicron detector.
The Harshaw resolution measurements conducted during this study are consistent with the historical measurements.

During operation, a single detector plus photomultiplier tube (PMT) is connected to an ORTEC digiBASE for digitization and readout, which itself is coupled to an Intel NUC running Ubuntu Linux.
A radiation source (either ${\sim}60$~{\textmu}Ci Eu-154 for calibration or ${\sim}7$~{\textmu}Ci Cs-137 for fuzzing experiments) is placed ${\sim}30$~cm away from one detector face, and the source and detector are partially enclosed by lead bricks to help isolate the setup from the radiation environment of the surrounding laboratory (and vice-versa).
This setup produces timestamped listmode gamma ray data with energies digitized into 1024 uncalibrated channels, and the following subsection describes the calibration procedures required for consistent comparisons of spectra across multiple detectors.

\begin{table}[!htbp]
\centering
\caption{NaI detectors used}
\label{tab:det-resolution}
\begin{tabular}{l|l|l|l}
\textbf{Detector ID:} & \textbf{010} & \textbf{238} & \textbf{269}\\\hline\hline
Manufacturer & Harshaw & Harshaw & Bicron\\\hline
Earliest tested date & Jan.~1998 & Feb.~1985 & unknown \\\hline
Resolution (then) [$\sigma$, $662$~keV] & 4.45\% & 3.87\% & unknown \\\hline
Resolution (now) [$\sigma$, $662$~keV] & 4.33\% & 3.80\% & 3.42\%
\end{tabular}
\end{table}

\subsection{Detector calibration}\label{sec:calibration}
The fuzz testing comparisons of Ref.~\cite{vavrek2025differential} operated on uncalibrated gamma spectra in order to compare detector behavior with a relatively low level of data postprocessing.
In this work, however, the natural variability in spectrum response to the same high voltage across different detectors would induce unacceptably large variations in the untampered baseline---in particular, the $\chi^2/\nu$ metric in Section~\ref{sec:fuzz_experiments}---potentially obscuring an attack signature.
As a result, we determine and apply an energy calibration for each detector to ensure valid comparisons.
While accurate channel-to-energy calibrations would depend on some of the fuzzed parameters (namely, the pulse width and fine gain), for simplicity we use only a single calibration per detector (at a fine gain of $0.8$ and a pulse width of $1.25$~{\textmu}s).
We also fix each detector's high voltage in advance based on qualitative previous tests, not necessarily using the same value for each detector.
As a result, in later fuzz testing at arbitrary fine gain and pulse width parameters, the ``calibrated'' peak energies will float somewhat, and the term ``apparent energy'' will be used as necessary.

\begin{figure*}[!htbp]
\centering
\begin{tikzpicture}
    \node[inner sep=0pt] (image_node) at (0,0) {
        \includegraphics[width=\linewidth]{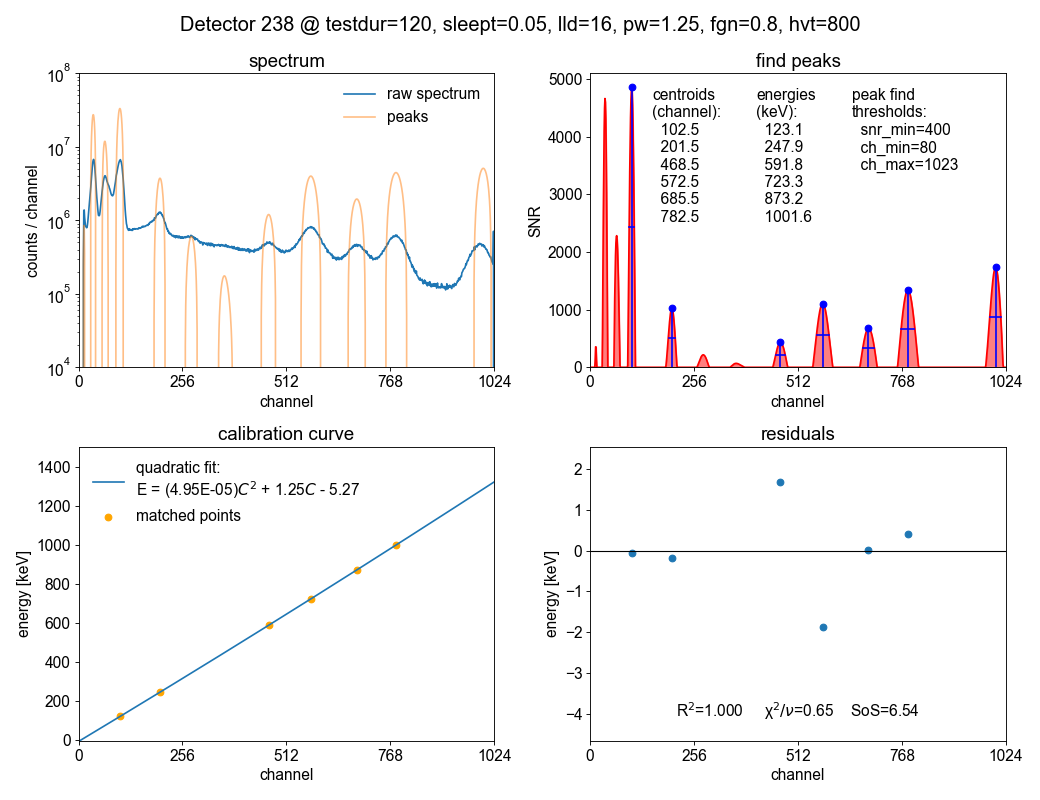}
    };
    \node[
        fill=white, 
        opacity=1.0, 
        text=black, 
        font=\sffamily,
        align=center,
        inner sep=5pt 
    ] at (0.5, 6.70) {
        \hspace{24pt}Detector 238 @ dwell time = 120 s, pw = 1.25, fgn = 0.8, hvt = 800\hspace{48pt}
    };
\end{tikzpicture}
\caption{
    Illustration of the calibration process for a single detector and parameter combination (detector 238, $120$~s dwell time, $1.25$~{\textmu}s pulse width, $0.8$ fine gain, $800$~V bias voltage).
    Top left and right: a convolution filter is applied to the raw spectrum to identify peaks.
    Bottom left: the identified peaks are matched to known energies of the Eu-154 source, and a quadratic fit is derived from the matched channel/energy pairs.
    Bottom right: quadratic fit residuals.
}
\label{fig:calibration}
\end{figure*}

\begin{figure}[!htbp]
\centering
\includegraphics[width=1.0\linewidth]{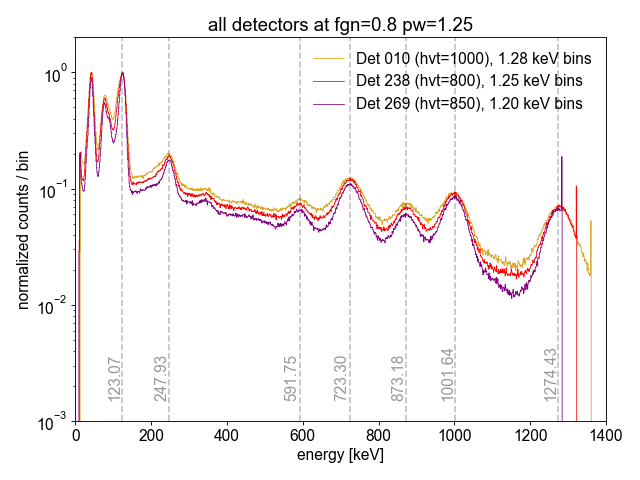}
\caption{
    Calibrated Eu-154 spectra from the three detectors at the given hvt, fgn, and pw settings.
    Each spectrum is normalized by its max bin content to better highlight differences in shape among the three detectors.
    Bin widths differ slightly due to different calibrations.
    The peak centroid at $1001.644$~keV is a weighted average of the $996.29$~keV and $1004.76$~keV lines with decay fractions of $0.1048$ and $0.1801$, respectively~\cite{nndc_eu154}.
    The last bin includes overflow.
}
\label{fig:calibrated-spectra}
\end{figure}

For each NaI detector, we derive a quadratic energy calibration to convert the uncalibrated channel numbers to photon energy depositions in keV, by mapping centroids of detected peaks from an Eu-154 measurement to known peak energies.
Fig.~\ref{fig:calibration} illustrates the steps of a single detector calibration, namely:
\begin{enumerate}
    \item We perform data acquisition with the detector to be calibrated to obtain the raw output spectrum (top left, blue).
    \item We apply a convolution filter to the raw spectrum to automatically detect peaks in the spectrum (top left, orange) and their associated signal-to-noise ratio (SNR) values (top right, red), using the {\tt becquerel} library~\cite{bandstra2023becquerel}.
    \item We manually find and apply a simple SNR threshold as well as min/max channel values to the peaks to select which ones are mapped to the Eu-154 peaks (top right, blue).
    This process is performed manually, but could potentially be done automatically.
    \item The most prominent photon energies of the Eu-154 spectrum are known from the literature~\cite{nndc_eu154} (bottom right).
    We map the previously-found peaks in the channel-binned spectrum to these energies to obtain 5-7 points on the calibration curve (bottom left, orange).
    We then perform a quadratic fit on these points to obtain the calibration function (bottom left, blue).
\end{enumerate}
The set of three consistently-calibrated detector spectra is shown in Fig.~\ref{fig:calibrated-spectra}.

\subsection{Fuzz testing experiments}\label{sec:fuzz_experiments}

\begin{figure}[!htbp]
\includegraphics[width=1.0\linewidth]{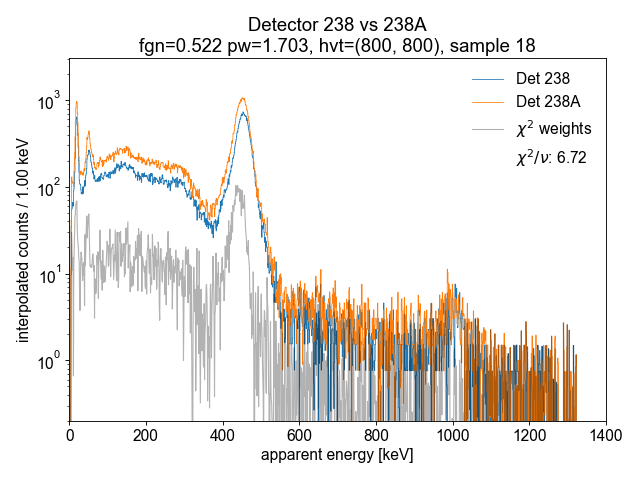}
\caption{
    Comparison of calibrated spectra collected with the same detector and parameters, with (orange, denoted 238A) and without (blue, denoted 238) the time-based attack triggering.
    After calibration, the spectra are interpolated to $1$-keV bins from $0$ to $1400$~keV, but, as described in Section~\ref{sec:calibration}, due to the calibration procedure, the apparent calibrated energies typically do not correspond to true energies.
    As such, the Cs-137 $662$~keV peak position appears closer to $400$~keV.
    The $\chi^2$ weights (gray), i.e., each of the $J$ summands in Eq.~\ref{eq:rchi2}, illustrate how much each spectrum bin contributes to the total $\chi^2/\nu$ statistic.
}
\label{fig:attack-example}
\end{figure}

We collected data from the three detectors using the same Cs-137 source.
Each detector dataset consists of $100$ $30$-second spectra taken over a fuzz sequence using the parameters in Table~\ref{tab:paramsweep}.
The fine gain (fgn), pulse width (pw), and system time values are sampled uniformly randomly from the given ranges for each test, using a random seed that is shared between all datasets to ensure that each data acquisition is done with the same fuzzing sequence.
For each detector, we performed two data acquisition sequences:
\begin{enumerate}
    \item A baseline set, where the detector is not subject to attack.
    \item An attacked set, where the detector is subject to the ``time-based attack'' used in Ref.~\cite{vavrek2025differential}.
        The demonstration attack shown here is active on one day of the week (arbitrarily chosen to be Sunday), therefore occurring with an overall frequency of $1/7$, and applies a duplication factor of $0.5$, i.e., gamma ray events observed by the detector have probability $P=0.5$ of being duplicated.
        Fig.~\ref{fig:attack-example} illustrates the effect of this attack on the detector output.
\end{enumerate}

\begin{table}[!htbp]
\centering
\caption{Parameter sweep values used in the fuzzing sequences}
\label{tab:paramsweep}
\setlength{\tabcolsep}{3pt}
\begin{tabular}{p{65pt}|p{80pt}|p{70pt}}

\textbf{Parameter}&
\textbf{Value}&
\textbf{Description} \\
\hline
\hline
Dwell time&
30~s&
Real-time measurement duration\\
\hline
High voltage (hvt)&
\makecell{
1000~V (det 010)\\
800~V (det 238)\\
850~V (det 269)}&
PMT bias voltage\\
\hline
Pulse width (pw)&
\makecell{Sampled randomly from\\
0.75 -- 2.0~{\textmu}s}&
Detector pulse width in microseconds\\
\hline
Fine gain (fgn)&
\makecell{Sampled randomly from\\
0.5 -- 1.2$\times$}&
Detector fine gain\\
\hline
System time&
\makecell{Sampled randomly from\\
7d ago -- present}&
\\
\end{tabular}
\end{table}

In this \textit{time-based attack} sequence, the host tampers with the detector measurements by using malicious code to selectively inject false duplicate data if the system time falls within some predetermined attack window.
This allows the host to remove (and secretly retain) half of the activity (which may be useful if the measured source is to be destroyed under the treaty) by simply duplicating listmode count data to try to fake the genuine signature.
Such an attack could be useful if the attacker knows that treaty verification measurements are to be conducted during a specific time period, though more advanced triggers are discussed in Ref.~\cite{vavrek2025differential}.
To detect such a time-based attack, the inspector can also include the system time in the fuzzed parameters in order to fool the attack into executing during testing.
In these experiments, we randomly select the system time from the interval between the real system time at the time of data collection and exactly seven days prior.
We also use the ``monkey-patching'' technique described in Ref.~\cite{vavrek2025differential} to tamper with the {\tt numpy.copy} array copy function implementation at runtime, such that inspection of the acquisition, analysis, and {\tt numpy} source code would not reveal the attack.

Comparison between two detector spectra $x$ and $y$ is performed via the modified reduced $\chi^2$ metric~\cite[Eq.~1]{vavrek2025differential},
\begin{align}\label{eq:rchi2}
    \chi^2 / \nu \equiv \frac{1}{J} \sum_{j=1}^{J} \frac{(x_j-y_j)^2}{x_j+y_j},
\end{align}
where $j = 1, \ldots , J$ is the spectrum bin index, we adopt the convention that $0 / 0 \equiv 0$, and we note that the metric is symmetric between $x$ and $y$.
As discussed in Ref.~\cite{vavrek2025differential}, other metrics may be chosen.
As a rule of thumb, values of $\chi^2/\nu \gg 1$ are said to indicate statistical inconsistency between the two spectra being compared, while values close to $0$ indicate less variation than expected by Gaussian noise.
In our previous work~\cite{vavrek2025differential} it was sufficient to set a fixed threshold of $2$ or $4$, depending on the attack, and alarm if any $\chi^2/\nu$ exceeded that value.
In Section~\ref{sec:results} of this work, however, we will find that (1) the absolute scale of non-attacked, multi-detector $\chi^2/\nu$ values be as large as ${\sim} 20$ in this measurement configuration; and (2) that the ``baseline'' of $\chi^2/\nu$ values may not be a flat line in certain useful visualizations.
As such, while we still use the $\chi^2/\nu$ metric throughout this work, its distance from $1$ is not on its own a useful measure.
Rather, it is used to establish an inter-detector $\chi^2/\nu$ signature that can be inspected by eye for deviations from a non-linear baseline (and could, in the future, be analyzed by more advanced anomaly detection methods).
Additionally, we note that the two spectra must have identical binning for the $\chi^2/\nu$ metric to be valid.
As a result of the calibration process, however, different detectors will end up with different energy bin widths and ranges.
We therefore (linearly) interpolate spectra to a common $1$~keV bin width from $0$ to $1400$~keV (zero padding where necessary) before computing the $\chi^2/\nu$ metric.
This interpolation procedure also changes the scale of the $\chi^2/\nu$ values, since the $J$ bins are no longer independent, so again interpreting the $\chi^2/\nu$ values must be done with respect to an initial baseline.

\section{Results and analysis}\label{sec:results}

\begin{figure}[!htbp]
\includegraphics[width=1.0\linewidth]{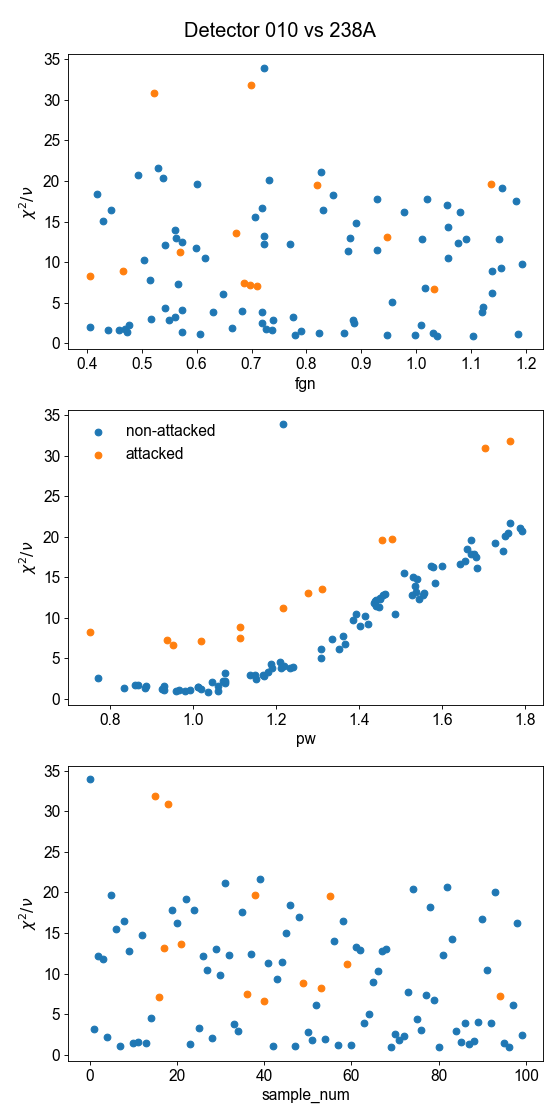}
\caption{
    An example of $\chi^2/\nu$ statistic distributions against fine gain (fgn, top), pulse width (pw, middle), and sample number (bottom) obtained from inter-detector fuzzing between detectors 010 and 238, with the latter subject to the time-based attack (suffix ``A'').
    Acquisitions in which the attack actually triggered are shown in orange while non-attacked samples are shown in blue.
}
\label{fig:chisq_vs_all}
\end{figure}

Recall that our objective is to identify the presence of a time-based attack on one detector given a second detector as reference.
We can establish the presence of an attack only if there is a clear relationship between the two detectors and the attacked data falls outside of that relationship.
Fig.~\ref{fig:chisq_vs_all} shows the relationships between detectors 010 and 238A in terms of the $\chi^2/\nu$ metric during fuzz testing; the ``A'' suffix denotes that the attack code was present for the set of acquisitions with the given detector, while the orange ``attacked'' points denote acquisitions where the attack actually triggered due to the fuzzed system time falling on a Sunday.
In contrast to the results of Ref.~\cite{vavrek2025differential}, where anomalies were clear when the $\chi^2/\nu$ was plotted against the sample number, here anomalies are difficult to detect in plots vs.\ sample number (or fine gain) but become clear when plotting against the pulse width (pw) setting.
This phenomenon was observed for all tested detector pairs.
The analysis vs.\ sample number or fine gain is also complicated by the high-$\chi^2/\nu$ artifact observed at sample~$0$, which appears to be related to device startup, and which was also observed in Ref.~\cite[Fig.~6]{vavrek2025differential}.
These sample-$0$ artifacts are occasionally but not always present, and are around the same magnitude in $\chi^2/\nu$ as the attacked samples despite (typically) not being attacked themselves.

\begin{figure}[!htbp]
\includegraphics[width=1.0\linewidth]{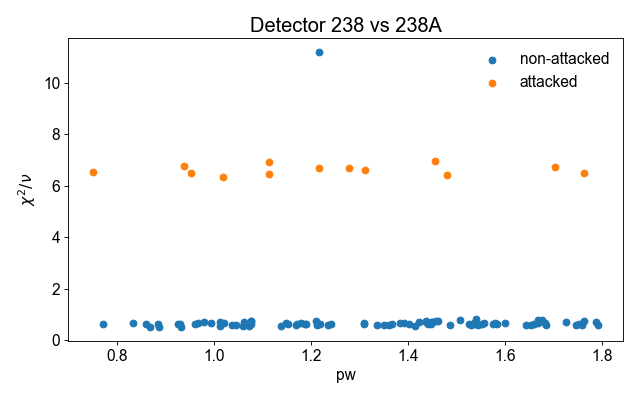}\\
\includegraphics[width=1.0\linewidth]{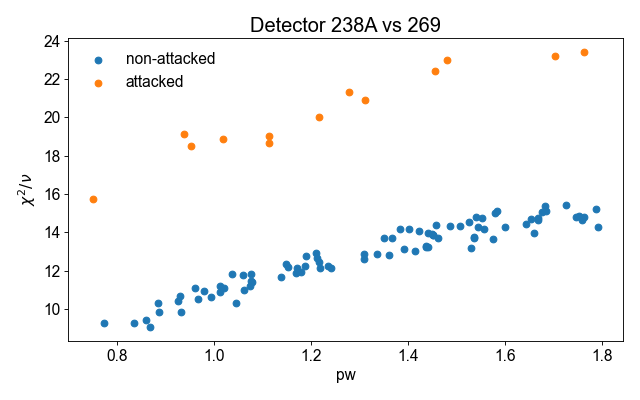}\\
\includegraphics[width=1.0\linewidth]{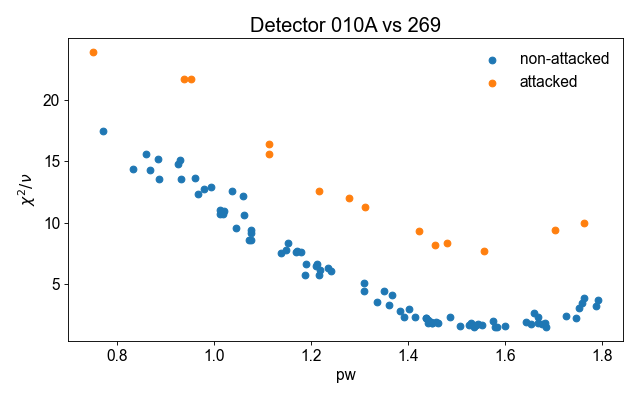}\\
\includegraphics[width=1.0\linewidth]{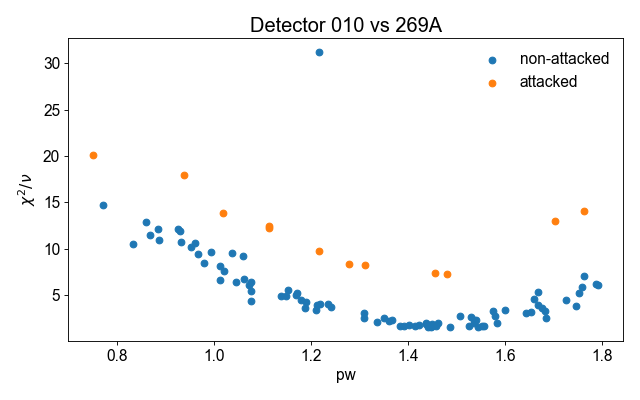}
\caption{
    Additional examples of the $\chi^2/\nu$ statistic vs.\ pulse width (pw) obtained from inter-detector fuzzing with various detector pairs (see also Fig.~\ref{fig:chisq_vs_all}, middle).
}
\label{fig:chisq_pw_only}
\end{figure}

Fig.~\ref{fig:chisq_pw_only} shows four additional examples of $\chi^2/\nu$ vs.\ pulse width.
Attacked and non-attacked samples are easily separable by a constant threshold when an attacked detector is compared against its own non-attacked baseline (238 vs.\ 238A).
Similar to Fig.~\ref{fig:chisq_vs_all}, the inter-detector comparisons (238A vs.\ 269, 010A vs.\ 269, and 010 vs.\ 269A) instead all exhibit a non-linear trend in $\chi^2/\nu$ vs.\ pulse width, with the attacked points sitting a roughly constant offset above this non-linear baseline.
Although the shape of the non-linear baseline varies among detector pairs, making an alarm threshold more difficult to define, the tamper attempt is clearly evident to the eye.
With the duplication fraction of $P = 0.5$ used here, the increase in $\chi^2/\nu$ resulting from the attack is typically much larger than the observed level of variation in the baseline.

Fig.~\ref{fig:chisq_pw_only} also shows that the attack can be detected on a Harshaw detector ($238$) dating to at least $1985$ based on the reference signature from the Bicron detector ($269$) from an unknown date, suggesting that the inter-detector fuzz testing method is robust to differences in manufacturer and to decades of detector aging.
Similarly, Figs.~\ref{fig:chisq_vs_all} and \ref{fig:chisq_pw_only} show that the attack on a better-resolution detector (238 or 269) can be detected using the baseline from a poorer-resolution detector (010), and vice-versa.

As mentioned in Section~\ref{sec:fuzz_experiments}, the $\chi^2/\nu$ metric values in Figs.~\ref{fig:chisq_vs_all} and \ref{fig:chisq_pw_only} should not be interpreted according to the na\"ive rule of ``$\chi^2/\nu \approx 1 \Leftrightarrow $ no attack.''
Most inter-detector comparisons here exhibit baseline $\chi^2/\nu$ values over a range of ${\sim} 3$--$15$ due to normal inter-detector and run-to-run differences alone, without the attack triggering.
It is the excess $\chi^2/\nu$ above these values that reveals the attack.
In addition, the same-detector comparison (238 vs.\ 238A) in Fig.~\ref{fig:chisq_pw_only} exhibits a non-attacked baseline of $\chi^2/\nu \approx 0.6$, very consistently less than $1$, which might na\"ively suggest that the spectra compared are ``too similar'' and thus also anomalous.
However, the consistently-low $\chi^2/\nu$ values here are likely the result of the post-calibration interpolation to the fixed binning, which reduces the statistical independence of adjacent bins.
Again, the attack is detected via anomalies from this baseline, regardless of the absolute scale of the baseline.

\section{Discussion}\label{sec:discussion}
The results of Section~\ref{sec:results} show that it is feasible to detect our representative tamper attempt using inter-detector fuzzing, even despite normal inter-detector differences.
Here we discuss some limitations and opportunities for future work.

First, the limitations identified in our original single-detector concept~\cite{vavrek2025differential} still apply.
For this inter-detector extension, it is worth re-emphasizing that environmental variations between the downrange and golden copy detectors can also increase $\chi^2/\nu$ values and could lead to false alarms if not well-controlled (and/or if an alarm threshold is set too low).

Second, our inter-detector proof-of-concept has been demonstrated on a small set of three detectors.
It would be valuable in the future to conduct additional multi-detector fuzz tests with a wider array of NaI detectors of various ages, manufacturers, and performance characteristics, in order to more fully explore the limits of inter-detector comparisons.
It would also be interesting to test whether inter-detector fuzz testing could be extended to include other digiBASE-compatible scintillator detectors, perhaps comparing ``medium-resolution'' LaBr$_3$ used in safeguards~\cite{iaea2022itvs} to low-resolution NaI.

As with the single-detector concept, and as mentioned in the Introduction, inter-detector fuzz testing is an anomaly \textit{detection} technique but not necessarily an anomaly \textit{characterization} technique.
Anomalous $\chi^2/\nu$ values are therefore only a starting point for further investigation (perhaps via the $\chi^2$ weights as in Fig.~\ref{fig:attack-example}), and could be indicative of accidental damage or degradation rather than malicious tampering.
Although we have shown that the technique is robust to variations and overall increases in $\chi^2/\nu$ from ``normal'' inter-detector differences, it would be interesting in future work to test the technique with damaged detectors and characterize the potential false positive anomalies.

We note again that visualizing the $\chi^2/\nu$ values vs.\ different fuzzed parameters can be crucial for distinguishing attacked samples from the non-attacked signature, and how the ``baseline'' may in fact not be a line.
In our previous work~\cite{vavrek2025differential}, it was sufficient to plot $\chi^2/\nu$ values vs.\ sample number and set a constant empirical threshold.
In inter-detector comparisons, more advanced anomaly detection methods may be required if the distributions are not so easily separable by eye.
Various standard clustering algorithms may be of use here.
Equipped with such anomaly detection methods, it would be interesting to quantify performance especially at smaller duplication fractions than the $P=0.5$ used for demonstration here.

\section{Conclusions}
We have extended the concept of physical differential fuzz testing, used to compare a radiation detector against its own previously-established baseline, to comparisons between multiple detectors.
The relationships between detector pairs can be consistent enough to enable tamper detection on a candidate detector via deviations from a baseline established on a ``golden copy'' reference detector, even among detectors with different manufacturers and performance characteristics.
This baseline is typically non-linear and may necessitate more advanced anomaly detection methods in the future.
Overall, this inter-detector fuzz testing concept can improve the operationalization of fuzz testing for radiation detectors in nuclear safeguards and treaty verification contexts, enabling a ``chain of knowledge'' between a trusted reference detector and in-field detectors where the baseline is incomplete or non-existent.

\section*{Acknowledgments}

The authors thank David Berger, Rebecca Blackmon, Jerome Meyer (US Defense Threat Reduction Agency) and LT Thomas Duane (US Navy) for the useful discussions that led to the idea of inter-detector fuzzing.
The authors also thank Kai Vetter and Ali Hanks (UC Berkeley) for loaning the NaI detectors, and Marco Salathe (LBNL) for assistance with data acquisition.

This document was prepared as an account of work sponsored by the United States Government.
While this document is believed to contain correct information, neither the United States Government nor any agency thereof, nor the Regents of the University of California, nor any of their employees, makes any warranty, express or implied, or assumes any legal responsibility for the accuracy, completeness, or usefulness of any information, apparatus, product, or process disclosed, or represents that its use would not infringe privately owned rights.
Reference herein to any specific commercial product, process, or service by its trade name, trademark, manufacturer, or otherwise, does not necessarily constitute or imply its endorsement, recommendation, or favoring by the United States Government or any agency thereof, or the Regents of the University of California.
The views and opinions of authors expressed herein do not necessarily state or reflect those of the United States Government or any agency thereof or the Regents of the University of California.

This manuscript has been authored by an author at Lawrence Berkeley National Laboratory under Contract No.~DE-AC02-05CH11231 with the U.S.~Department of Energy.
The U.S.~Government retains, and the publisher, by accepting the article for publication, acknowledges, that the U.S.~Government retains a non-exclusive, paid-up, irrevocable, world-wide license to publish or reproduce the published form of this manuscript, or allow others to do so, for U.S.~Government purposes.

\bibliographystyle{IEEEtran}
\bibliography{biblio}

\end{document}